# Consistent Initialization of the Laplace Transform

S. Ahuja[a*] and R. K. Arya[b]

[a*]Corresponding Author
Department of Chemical Engineering,
Thapar Institute of Engineering & Technology,
Bhadson Road,
Patiala-147004, Punjab, India. E-mail: skahuja@thapar.edu

[b]Department of Chemical Engineering,
National Institute of Technology,
Grand Trunk Road, Bye pass,
Jalandhar-144011, Punjab, India. E-mail: aryark@nitj.ac.in

*Abstract*–**Consistent initialization of the Laplace transform has been a fundamental and long-standing issue. The consistency of the $\mathcal{L}_-$ approach has been questioned, yet it is a popular approach since the $\mathcal{L}_+$ approach requires a priori computation of the $0^+$ initial conditions, which becomes a diligent task from the available methods. Also, the $\mathcal{L}_+$ Laplace transform of the impulse becomes zero, thus involving an inconsistency. In contrast to these direct approaches, some studies propose convoluted methods to address the issue. Here, a direct, facile, and first-principles novel approach for the $\mathcal{L}_+$ transform is proposed. It computes the consistent $0^+$ initial conditions and solution based on singular-nonsingular decomposition of the system model. The inconsistency associated with the $\mathcal{L}_+$ Laplace transform of discontinuous functions and impulse is not involved in the nonsingular part. The emergence of discontinuity is easily elucidated from the singular part, and the solution yields the same $0^+$ initial values as the initial conditions used. Further, physical first-principles can be used in the midst of the computation to validate the results to ensure error-free execution.**

## I. INTRODUCTION

### A. Background

The unilateral Laplace transform is widely used to treat time-invariant systems for inputs with jump discontinuities such as impulse and step. These find extensive applications in diverse fields [1]-[3]. Impulse and its derivatives are called singularities since these are not well-defined mathematical functions. A significant issue in determining their dynamic response is the ability to mathematically treat discontinuity and transients at the origin. The transients may include the non-zero and unsteady initial state, and singularity emerging from the differential of a discontinuous term in the model of a process. A very short duration change in an operating variable can be simulated to an impulse.

There has been a considerable resurgent interest on the long-standing issue of inconsistency in the Laplace transform in the literature [3]-[14]. Many literature, for example [15]-[17], define the right-sided Laplace transform $F(s)$ of a function $f(t)$ and the corresponding derivative rule as

$$F(s) = \mathcal{L} f(t) = \int_0^\infty f(t)e^{-st}dt, \quad (1)$$

$$\mathcal{L} f'(t) = sF(s) - f(0). \quad (2)$$

However, they do not clarify where the integration should start and what is $f(0)$ exactly if $f(t)$ has a discontinuity at the origin. Some literature [7],[12],[18]-[20] support the use of the $\mathcal{L}_-$ Laplace transform, which takes $0^-$ in the definition of the Laplace transform and its derivative rule as

$$F(s) = \mathcal{L}_- f(t) = \int_{0^-}^\infty f(t)e^{-st}dt, \quad (3)$$

$$\mathcal{L}_- f'(t) = sF(s) - f(0^-). \quad (4)$$

The $\mathcal{L}_-$ approach directly uses the available $0^-$ or pre-initial conditions just before the input.

The $\mathcal{L}_+$ approach advocated in literature [3],[11],[21]-[23], however, is a two-step solution. In the first step, the $0^+$ or the post-initial conditions just after the input are commonly calculated using physical reasoning. These are then used in the $\mathcal{L}_+$ transform and its derivative rule defined as

$$F(s) = \mathcal{L}_+ f(t) = \int_{0^+}^\infty f(t)e^{-st}dt, \quad (5)$$

$$\mathcal{L}_+ f'(t) = sF(s) - f(0^+). \quad (6)$$

The $\mathcal{L}_-$ approach is a popular approach since it avoids a priori calculation of the post-initial conditions unlike the $\mathcal{L}_+$ approach [24]. However, its consistency has been questioned in the literature [3],[8],[9]. The inconsistency arising essentially due to the $f(0)$ term of the derivative rule during the solution, has been circumvented in various possible ways.

The derivative rule (4) of the $\mathcal{L}_-$ approach uses pre-initial $f(0^-)$ conditions. In the case of discontinuity, the initial values obtained from the solution on $t \geq 0$ are not the same as the original $f(0^-)$ conditions used for the solution [8]. However, it is naturally expected of the solution to satisfy the initial conditions that were originally used to initialize the treatment of the differential equation. To resolve this, the superposition of the responses based on the zero initial conditions and zero input function has been presented [8]. Alternatively, the original differential equation can be transformed by adding on its right certain terms having delta and its derivatives, thus converting it into a generalized functions (distributions) form. The resulting equation can then be solved for the zero initial conditions again using the derivative rule $\mathcal{L} f'(t) = sF(s)$ to get the final evolution [9]. Another study on the other hand avoids the advanced machinery of generalized functions [10]. However, it involves complicated transformation of the original differential equation in turn. New continuous functions are then defined that are the integrals of the original functions and thus do not involve the discontinuous $f(0)$ term. It can, however, be shown that the strategy does not work for the impulse perturbation case.

In contrast to the aforesaid indirect methods that require additional mathematical proofs [8]-[10], the $\mathcal{L}_-$ and $\mathcal{L}_+$ are direct approaches to solution requiring neither transformations nor superposition. However, in the $\mathcal{L}_+$ approach, the Laplace transform of the unit-impulse or Dirac delta becomes zero from the derivative property (6), [7]

$$\mathcal{L}_+\{\delta(t)\} = s\mathcal{L}\theta(t) - \theta(0^+) = s\frac{1}{s} - 1 = 0, \qquad (7)$$

since the unit-impulse function $\delta(t)$ is the derivative of the unit-step function $\theta(t)$, thus causing an inconsistency. It has, hence, been claimed that the $\mathcal{L}_+$ approach generates identically zero solution for certain cases of differentiated discontinuous inputs or delta function in the model [12]. However, it can be shown that this claim is erroneous. In fact, such inputs cause initial jump discontinuity of the state, and so first the relevant $0^+$ initial conditions of the state need to be established. Henceforth, the $\mathcal{L}_+$ transformation during the second step yields the correct solution. Besides, it has also been stressed to appropriately specify the pre-initial value of the differentiated input that has a discontinuity depending upon a physical situation [13].

The impetus for the $\mathcal{L}_+$ approach also comes in from a recent work [3] demonstrating the possible pitfalls of the above purely model-based approaches [8]-[10] including the $\mathcal{L}_-$ approach. An approach involving physical principles can yield more accurate post-initial conditions and, hence, the solution than any purely mathematical approach. This, in fact, can occur for the impulse perturbation cases of certain linear and nonlinear nontrivial applications [2]-[6]; notwithstanding the fact that well-established models from the standard literature have been employed, and the same models exhibit no such issue for the step perturbation case. Thus, indicating that the results are not due to the inadequacy of the mathematical formulation or specification of the models. For instance, consider an impulse perturbation introduced into the inflow rate to a gravity-flow tank containing a tank and exit pipe [3],[6]. This perturbation can be physically realized by plunging an additional quantity of liquid into the tank. The $\mathcal{L}_-$ or purely mathematical approaches render zero initial discontinuity in the velocity of efflux off the exit pipe [3],[6]. However, this result is not physically validated since there is a sudden level change that causes a simultaneous change in the efflux velocity. Some consistency issues thus emerge during application besides the mathematical issues.

The different approaches to consistent initialization presented in the literature are in general difficult to apply. These involve mathematical technicalities and transformations of models, and cannot be easily implemented on a routine basis by the researchers and professionals of various applied fields. The challenge thus lies in circumventing the mathematical complexity by achieving consistent initialization and solution through direct methods. The $\mathcal{L}_-$ and $\mathcal{L}_+$ are, however, direct approaches to solution. Unfortunately, in either approach, the passage across $t = 0$ is a potential source of error. The present article attempts to provide principles for passing from $0^-$ to $0^+$ initial conditions, and subsequent solution by means of the Laplace analysis.

### B. Consistent Initialization and Analysis

The above discussion motivates us to use the $\mathcal{L}_+$ approach for a direct, facile, and consistent right-sided Laplace treatment. For the systems of practical interest the computed solution should be able to predict the evolution for the non-negative half-line. Consistent with this the $f(0^+)$ post-initial conditions used in the $\mathcal{L}_+$ approach are also supported on the non-negative half-line.

A difficulty of the $\mathcal{L}_+$ approach, however, is that it requires additional effort of applying physical principles to compute the $0^+$ initial conditions before carrying out the solution. The task of having a physical insight into the system's discontinuous behavior is case-specific, and requires user's diligence. In an impulse matching technique the input and output function terms of the highest order of their derivatives in the model are matched [7]. This technique of computing initial conditions again becomes difficult for higher-order systems [7]. Another approach presented [10] is also based on a convoluted technique noted above.

A modified form of the $\mathcal{L}_+$ approach is proposed here. It is a three-stage methodology that establishes consistent post-initial conditions and solution. The methodology is direct, facile, first-principles, and is not case-specific.

## II. PROPOSED METHODOLOGY

The proposed methodology is based on singular-nonsingular decomposition of the model represented by generalized functions. The given system equation is decomposed into its singular and nonsingular parts that are solved separately. In the first stage, the solution to the singular part is obtained. Secondly, the post-initial conditions are computed from the singular part of the solution. Discontinuity analysis is carried out to find out which variables of the state jump to a new value and by how much. Thirdly, the regular part of the solution is generated from the Laplace transform. The obtained post-initial conditions successively initialize the solution for the computation of the evolution of the system. The singular and nonsingular parts are eventually combined to yield the final solution.

A linear time-invariant model of a system is considered on $[0, \infty)$ represented by the following linear ordinary differential equation with constant coefficients

$$y^{(n)}(t) + a_1 y^{(n-1)}(t) + \ldots a_n y(t) = b_0 x^{(m)}(t) + b_1 x^{(m-1)}(t) + \ldots b_m x(t), \quad (8)$$

where $y^{(n)}(t)$ denotes the $n^{th}$ derivative of $y$, etc. $y^{(0)}(t) = y(t)$ in particular. Here, $y$ is an output function to be solved and $x$ is a known input function, for example, a step or impulse occurring at time $t = 0$; $a_1 \ldots a_n$ and $b_0, b_1 \ldots b_m$ are real coefficients, and $n \geq m \geq 0$. The methodology employs generalized functions. An input/output function $f$ is composed of regular and singular parts $f_r$ and $f_s$, respectively. The regular part of the function is piecewise smooth on $[0, \infty)$, such that all its derivatives exist on the complement of a sparse set of numbers [7]. Its left and right limits exist at the origin and at each element of the sparse set. Left limit at the origin stands for the pre-initial value of the piecewise smooth function and its derivatives. The singular part of the generalized function contains the singularity functions on $[0, \infty)$ at the isolated points of the sparse set [7].

Since Eq. (8) must be balanced at any time $t$, it can be written for both the singular and the regular parts of the functions. Hence, the system can be decomposed as

$$y_s^{(n)}(t) + a_1 y_s^{(n-1)}(t) + \ldots a_n y_s(t) = b_0 y_s^{(m)}(t) + b_1 y_s^{(m-1)}(t) + \ldots b_m y_s(t), \quad (9)$$

$$y_r^{(n)}(t) + a_1 y_r^{(n-1)}(t) + \ldots a_n y_r(t) = b_0 x_r^{(m)}(t) + b_1 x_r^{(m-1)}(t) + \ldots b_m x_r(t). \quad (10)$$

### A. Singular Part of the Solution

The setting for the first stage of the solution is made. After placing the known input, successive integration of Eq. (8) would result in a system of differential equations of decreasing consecutive orders. A constant of integration is introduced at each step, thus

$$y^{(n)}(t) + a_1 y^{(n-1)}(t) + \ldots a_n y(t) = b_0 x^{(m)}(t) + b_1 x^{(m-1)}(t) + \ldots b_m x(t),$$

$$y^{(n-1)}(t) + a_1 y^{(n-2)}(t) + \ldots a_n y^{(-1)}(t) = b_0 x^{(m-1)}(t) + b_1 x^{(m-2)}(t) + \ldots b_m x^{(-1)}(t) + c_1,$$

$$y^{(n-2)}(t) + a_1 y^{(n-3)}(t) + \ldots a_n y^{(-2)}(t) = b_0 x^{(m-2)}(t) + b_1 x^{(m-3)}(t) + \ldots b_m x^{(-2)}(t) + c_1 t + c_2,$$

$$\ldots$$
$$\ldots$$

$$y^{(1)}(t) + a_1 y^{(0)}(t) + \ldots a_n y^{(-n+1)}(t) = b_0 x^{(m-n+1)}(t) + b_1 x^{(m-n)}(t) + \ldots b_m x^{(-n+1)}(t) + g'(t),$$

$$y^{(0)}(t) + a_1 y^{(-1)}(t) + \ldots a_n y^{(-n)}(t) = b_0 x^{(m-n)}(t) + b_1 x^{(m-n-1)}(t) + \ldots b_m x^{(-n)}(t) + g(t), \quad (11)$$

where $c_1, c_2$, etc. are constants, $g(t)$ is a polynomial in $t$, and $y^{(-n)}$ is the $n^{th}$ integral of $y$. Thus, for $n=1$, $y^{(-1)} = \int_0^t y(t) dt$.

The singular parts of $y^{(n)}, y^{(n-1)}$, etc. can be computed from the set of Eqs. (11). It can be seen in the first equation of this set that $x$ and $y$ become increasingly singular as the order of their differentiation increases. For an input impulse in $x(t)$ at $t = 0$, the term of the highest singularity on the right of this equation would be $b_o \delta^{(m)}(t)$. The term producing this singularity on the left is the term of the highest singularity, $y_s^{(n)}(t)$. Thus, the maximum singularity in $y(t)$ is, $y_s(t) = b_o \delta^{(m-n)}(t)$. Further, since $n \geq m \geq 0$, the maximum possible singularity for $y(t)$ would occur when $n = m$, so, $y_s(t) = b_o \delta^{(m-n)}(t) = b_o \delta^{(0)}(t) = b_o \delta(t)$. From this reflection, $y^{(-1)}, y^{(-2)} \ldots, y^{(-n)}$ and $x^{(-1)}, x^{(-2)} \ldots, x^{(-n)}$ appearing in the successive equations of (11) must not contain any singularity. Also, the regular polynomials in $t$ have no singularity. Consequently, the corresponding singular parts of Eqs. (11) become,

$$y_s^{(n)}(t) + a_1 y_s^{(n-1)}(t) + \ldots a_n y_s(t) = b_0 x_s^{(m)}(t) + b_1 x_s^{(m-1)}(t) + \ldots b_m x_s(t),$$

$$y_s^{(n-1)}(t) + a_1 y_s^{(n-2)}(t) + \ldots a_{n-1} y_s(t) + 0 = b_0 x_s^{(m-1)}(t) + b_1 x_s^{(m-2)}(t) + \ldots b_{m-1} x_s(t) + 0 + 0,$$

$$y_s^{(n-2)}(t) + \ldots a_{n-2} y_s(t) + 0 + 0 = b_0 x_s^{(m-2)}(t) + b_1 x_s^{(m-3)}(t) + \ldots + b_{m-2} x_s(t) + 0 + 0 + 0,$$

$$\ldots$$
$$\ldots$$

$$y_s^{(1)}(t) + a_1 y_s(t) + 0 \ldots 0 = b_0 x_s^{(m-n+1)}(t) + b_1 x_s^{(m-n)}(t) + 0 \ldots 0,$$

$$y_s(t) + 0 \ldots 0 = b_0 x_s^{(m-n)}(t) + 0 \ldots 0. \quad (12)$$

The last equation of the set of Eqs. (12) yields the solution $y_s(t)$ for the known input $x(t)$. This solution in turn can be placed into the just preceding equation to give $y_s^{(1)}$. Proceeding backward similarly, the set of Eqs. (12) are solved simultaneously in an algebraic manner to yield the singular parts of solution for $y_s^{(2)}, \ldots, y_s^{(n)}$, respectively.

### B. Post-initial Conditions

In the second stage, the above singular parts can be used to compute the initial discontinuity values. In general, the singular part of the differential $y^{(n)}(t)$, that is $y_s^{(n)}(t)$, $n = 1, \ldots, n$, arises from the jump discontinuity in the values of $y^{(n-1)}(t)$ at zero time, that is at $a = 0$ in the following equation, [7]

$$y_s^{(n)}(t) = (y^{(n-1)}(a^+) - y^{(n-1)}(a^-))\delta(t-a) = (y^{(n-1)}(0^+) - y^{(n-1)}(0^-))\delta(t-0). \quad (13)$$

If Eq. (13) is integrated it yields the initial jump discontinuity values in $y^{(n-1)}(t)$, $n = 1, \ldots, n$, since the integral of the Dirac delta function is one. Thus,

$$\int_0^\infty y_s^{(n)}(t)dt = y^{(n-1)}(0^+) - y^{(n-1)}(0^-). \qquad (14)$$

Hence, the initial jump discontinuity of the state $y^{(n-1)}(t)$ is contributed only by the integral of the singular part of its derivative $y_s^{(n)}(t)$. The integral of the regular part $y_r^{(n)}(t)$ cannot contribute anything to the jump discontinuity due to the absence of singularity in $y_r^{(n)}(t)$. Thus, the singular parts computed above in the first stage are integrated from 0 to ∞ to generate the initial discontinuity and so the post-initial condition for the derivatives of the output, $y^{(n-1)}(0^+)$, $n = 1, \ldots, n$.

*C. Physical Validation*

For validation of the above post-initial conditions, an alternative method can be used. The post-initial conditions are determined by the direct application of the physical first-principles of mass, energy, and momentum balances to a system at the initial instant of the introduction of the input. These principles are instinctive and, hence, less prone to errors. Thus, they can be used to check the correctness of the results for most common cases. Further, as noted in the introduction the physical principles can lead to more accurate results than a model-based approach for certain cases. The results from the two methods can be compared to ensure error-free execution. Once the consistent initial conditions are established, these can be used to compute the solution from the regular part in the third stage described next.

*D. Regular Part of the Solution*

In the third stage the solution is carried out by means of the $\mathcal{L}_+$ Laplace transform. In the conventional $\mathcal{L}_+$ approach, the original model Eq. (8) containing generalized functions with singular and nonsingular parts combined is employed. However, in the present work, Eq. (10) representing the regular part of the model is proposed to be used for the Laplace transformation. The singularity functions and discontinuity are thus circumvented. The above post-initial conditions are used to initialize the transformed equation. The solution for the regular part $y_r(t)$ is obtained in the usual manner after inversion. This solution is combined with the singular part $y_s(t)$ obtained above to give the generalized solution to the system, finally

$$y(t) = y_r(t) + y_s(t). \qquad (15)$$

### III. EXAMPLE

The U-tube manometer system shown in Fig. 1 measures a pressure difference $p$ imposed across its two legs. It contains a liquid of density $\rho$ and viscosity $\mu$. For the liquid column, $v$ is its velocity, $m$ is its mass, $L$ is its total linear length, and $A$ is its cross-sectional area. The acceleration due to gravity is $g$. Assuming laminar flow in the tube and flat velocity profile for the inertial terms, the momentum balance gives the following second-order linear ordinary differential equation with constant coefficients for $v$ that contains the derivative of the input $p$.

$$m\ddot{v}(t) + l\dot{v}(t) + kv(t) = A\dot{p}(t), \quad t \geq 0, \qquad (16)$$

where constants $l = 8\mu LA$, and $k = 2A\rho g$. The pre-initial conditions $v(0^-)$, $\dot{v}(0^-)$, and $p(0^-)$ have known values.

Considering impulse input in the applied pressure difference as

$$p(t) = p(0^-) + M\delta(t), \quad t \geq 0, \qquad (17)$$

where $M$ is the magnitude of perturbation and $\delta(t)$ is the Dirac delta function.

The proposed methodology is used. In the first stage, the singular parts of Eq. (16) and its successive integrals after placing the input signal as Eqs. (12) are

$$m\ddot{v}_s(t) + l\dot{v}_s(t) + kv_s(t) = A\dot{p}_s(t) = AM\dot{\delta}(t), \qquad (18)$$

$$m\dot{v}_s(t) + lv_s(t) + 0 = Ap_s(t) = AM\delta(t), \qquad (19)$$

$$mv_s(t) + 0 = 0. \qquad (20)$$

Solving these equations in the backward manner as described above, yields

$$v_s(t) = 0, \qquad (21)$$

$$\dot{v}_s(t) = \frac{AM}{m}\delta(t), \qquad (22)$$

$$m\ddot{v}_s(t) = -l\frac{AM}{m}\delta(t) + AM\dot{\delta}(t). \qquad (23)$$

In the second stage of the proposed methodology, using the integration of Eqs. (22) and (23) as in Eq. (14) for computing the post-initial conditions, we get,

$$\int_0^\infty \dot{v}_s(t)dt = v(0^+) - v(0^-) = \frac{AM}{m}, \qquad (24)$$

$$\int_0^\infty \ddot{v}_s(t)dt = \dot{v}(0^+) - \dot{v}(0^-) = -l\frac{AM}{m^2} + \frac{AM}{m}\delta(t) = -l\frac{AM}{m^2}. \qquad (25)$$

Eqs. (24) and (25) reveal the jump discontinuity and the post-initial condition of the velocity and acceleration of the column.

*Physical Validation* The initial conditions obtained from the momentum balance also happen to be the same. 'A' times the magnitude of the applied impulse equals the simultaneous change in the initial momentum of the column. Hence, the jump discontinuity in velocity is 'AM' divided by mass 'm'. This yields Eq. (24). Also, 'm' times the jump discontinuity in the initial acceleration of the liquid column is equal to the initial change in the frictional force offered due to the discontinuity in the initial velocity (term with coefficient 'l'). The initial change in the restoring force offered (term with coefficient 'k') is zero since the initial discontinuity in the displacement is zero. This eventually leads to the same Eq. (25).

In the third stage of the proposed methodology, the solution by means of the Laplace analysis is performed. Here, the regular part of Eq. (16) is used in place of the original Eq. (16) that is used in the conventional Laplace transform treatment. The advantage of doing so is discussed in the next section. The regular parts of the impulse input terms on the right vanish, leading us to

$$m\ddot{v}_r(t) + l\dot{v}_r(t) + kv_r(t) = A(\dot{p}(0^-) + (p(0^+) - p(0^-)))\delta'(t) = 0. \quad (26)$$

In the $\mathcal{L}_+$ transform treatment, Eq. (26) is solved using the post-initial conditions obtained above. Eq. (26) is $\mathcal{L}_+$ Laplace transformed as

$$m\{s^2V_r(s) - sv(0^+) - \dot{v}(0^+)\} + l\{sV_r(s) - v(0^+)\} + kV_r(s) = 0. \quad (27)$$

Taking the parameter values as $m = 1$ kg, $l = 2$ kg s$^{-1}$, $k = 1$ kg s$^{-2}$, and the pre-initial conditions as $v(0^-) = AM$ m s$^{-1}$ and $\dot{v}(0^-) = -2AM$ m s$^{-2}$. The post-initial conditions from Eqs. (24), and (25) become, $v(0^+) = 2AM$ m s$^{-1}$, and $\dot{v}(0^+) = -4AM$ m s$^{-2}$.

Placing all the parameter values into Eq. (27) leads to

$$1\{s^2V_r(s) - s(2AM) - (-4AM)\} + 2\{sv_r(s) - (2AM)\} - 1.v_r(s) = 0. \quad (28)$$

Thus, all the singular terms in Eq. (28) vanish, yielding

$$V_r(s) = \frac{2AM\,s}{(s+1)^2} = 2AM\left\{\frac{1}{(s+1)} - \frac{1}{(s+1)^2}\right\}, \quad (29)$$

and the solution for the regular part can be obtained through inversion as

$$v_r(t) = 2AM(e^{-t} - te^{-t}) \text{ m s}^{-1}, \ t \geq 0. \quad (30)$$

The generalized solution for the evolution of velocity of the column using Eqs. (21) and (30), is

$$v(t) = v_r(t) + v_s(t) = 2AM(e^{-t} - te^{-t}) \text{ m s}^{-1}, \ t \geq 0. \quad (31)$$

The evolution of column velocity is given in Fig. (2). The post-initial value can be verified from Eq. (29) using the initial value theorem as

$$v(0^+) = \lim_{s \to \infty} sV_r(s) = \lim_{s \to \infty} \frac{2AM\,s^2}{(s+1)^2} = 2AM, \text{ ms}^{-1}. \quad (32)$$

which is the same as the post-initial condition used for the solution above.

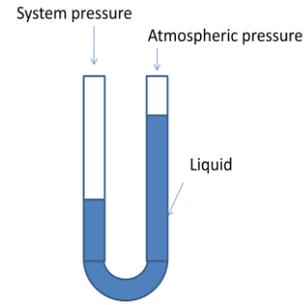

**Figure 1.** Schematic of a U-tube manometer system.

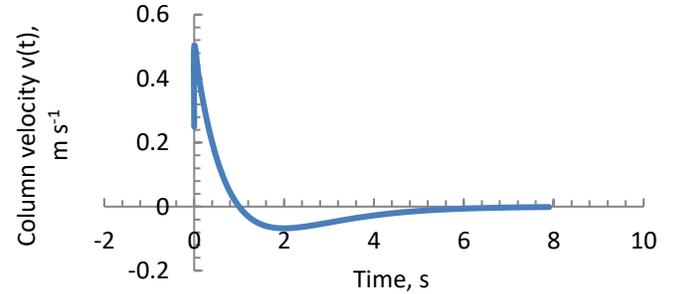

**Figure 2.** Evolution for the U-tube manometer system.

## IV. DISCUSSION

The methodology proposed here for the $\mathcal{L}_+$ approach is facile since it segregates singular and nonsingular parts of the solution. The first stage involves only the singular part of the differential equation and its solution, which successively establishes the relevant consistent initial conditions in the second stage. Therefore, it is easily elucidated how the jump discontinuity of the state originates due to the singularity of the input signal.

The third stage of the proposed methodology deals only with the nonsingular part of the system equation, so the $\mathcal{L}_+$ Laplace transformation only involves functions without discontinuity and singularity at the origin. This is in contrast to the conventional $\mathcal{L}_+$ approach in which the Laplace transformation of the original system equation is carried out. Thus, the inconsistency, that the $\mathcal{L}_+$ Laplace transform of the delta distribution is zero, is not involved in the proposed modified $\mathcal{L}_+$ approach. In the one-step $\mathcal{L}_-$ approach, however, the origin of the discontinuity is not clearly distinguishable since the different solution stages get clubbed. The implementation of the $\mathcal{L}_-$ transform also requires that any singularity functions with support at the origin must be included.

The application of the proposed methodology also demonstrates the consistency of the right-sided $\mathcal{L}_+$ approach for the initial conditions chosen. Unlike the $\mathcal{L}_-$ approach, it becomes straightforward for a user to identify that the initial values obtained from the solution on $t \geq 0$ here are consistent with the original $0^+$ conditions chosen for the solution. The literature studies propose transformed and convoluted methods to address the consistent initialization issues discussed in the introduction [8]-[10].

In the second stage of the proposed methodology, discontinuity analysis is employed for computing consistent post-initial conditions. It is a uniform and convenient method that can be used for a system of any order, for discontinuous or singular inputs, and for non-zero pre-initial conditions. Hence, it proves to be a clear and reliable alternative to the other available methods of computing post-initial conditions [7],[10]. Against this, the commonly used physical balances in the calculations of post-initial conditions for the $\mathcal{L}_+$ approach requires the application of physical laws such as the conservation of charge, mass, momentum, etc. to the system dynamics. This task of having a physical insight on the system's initially discontinuous behavior becomes difficult to a fresh user who initially has too little experience and diligence to have this. Furthermore, the investigation needs to be done for each system from the beginning since the physical reasoning is case-dependent. Nevertheless, physical principles can ensure error-free execution by generating precise post-initial conditions, and, hence, the subsequent solution as noted in the introduction. Consequently, these can be used for validating the mathematical results. Invoking physical principles in the midst of a computation and making it a step-wise process, reinforces these principles for fresh users, and is instinctive and so less susceptible to error for researchers and professionals.

*Acknowledgement* The authors are thankful to Ms Jyoti Sharma, Junior Research Fellow at the Department of Chemical Engineering, Thapar Institute of Engineering & Technology, Patiala, India for her help in the manuscript preparation.